\documentclass[aps, pra, superscriptaddress, reprint, floatfix]{revtex4-2}

\usepackage{graphicx}   
\usepackage[colorlinks = true, linkcolor=teal, citecolor=teal, urlcolor  = teal]{hyperref}
\usepackage{algorithm}  
\usepackage{algpseudocode}  
\usepackage{amsmath}    
\usepackage{amssymb}    
\usepackage{color}
\usepackage{physics}
\usepackage{tikz}
\usepackage{dsfont}
\usepackage{booktabs}
\usepackage{times}
\usepackage{comment}

\usepackage[capitalize]{cleveref}
\usepackage{quantikz}
\usepackage{svg}

\usetikzlibrary{shapes.geometric}
\usepackage{relsize}
\usepackage{tcolorbox}
\usepackage{enumerate}
\usepackage{amsthm}
\usepackage{orcidlink}

\crefname{equation}{Eq.}{equations}
\crefname{section}{Sec.}{sections}
\crefname{figure}{Fig.}{figures}
\crefname{appendix}{Appendix}{appendices}
\crefname{table}{Table}{tables}

\theoremstyle{plain}

\AfterEndEnvironment{theorem}{\noindent\ignorespaces}
\AfterEndEnvironment{lemma}{\noindent\ignorespaces}
\AfterEndEnvironment{corollary}{\noindent\ignorespaces}

\AfterEndEnvironment{definition}{\noindent\ignorespaces}
\AfterEndEnvironment{proof}{\noindent\ignorespaces}

\begin{document}

\title{Computational regimes in matrix-product-state-based quantum trajectory simulations}

\author{Aaron Sander\,\orcidlink{0009-0007-9166-6113}}
\affiliation{Technical University of Munich, Munich, Germany}

\author{Simon Cichy\,\orcidlink{0000-0002-9409-193X}}
\affiliation{Freie Universität Berlin, Berlin, Germany}

\author{Martin Eigel\,\orcidlink{0000-0003-2687-4497}}
\affiliation{Weierstrass Institute for Applied Analysis and Stochastics, Berlin, Germany}

\author{Jens Eisert\,\orcidlink{0000-0003-3033-1292}}
\affiliation{Freie Universität Berlin, Berlin, Germany}
\affiliation{Helmholtz-Zentrum Berlin für Materialien und Energie, Berlin, Germany}

\author{Maximilian Fröhlich\,\orcidlink{0009-0007-5276-2858}}
\affiliation{Weierstrass Institute for Applied Analysis and Stochastics, Berlin, Germany}

\author{Tom Peham\,\orcidlink{0000-0003-3434-7881}}
\affiliation{Technical University of Munich, Munich, Germany}

\author{Robert Wille\,\orcidlink{0000-0002-4993-7860}}
\affiliation{Technical University of Munich, Munich, Germany}
\affiliation{Munich Quantum Software Company GmbH, Munich, Germany}
\affiliation{Software Competence Center Hagenberg GmbH (SCCH), Hagenberg, Austria}

\date{\today}

\begin{abstract}
Efficient simulation of open quantum systems is central to modeling noisy quantum hardware and many-body
dynamics.
In trajectory-based tensor network methods, cost is often associated with trajectory-level quantities such as
entanglement growth or bond dimension.
However, the total cost of a fixed-accuracy simulation also depends on statistical sampling and on how many
trajectories can run in parallel.
Here we characterize computational regimes in matrix product state (MPS)-based quantum trajectory simulations
by separating memory per trajectory, runtime per trajectory, and sampling effort.
Two dimensionless ratios compare the bond dimensions and trajectory counts required by physically equivalent
unravelings to reach the same target accuracy.
When memory limits parallel execution, a smaller trajectory representation reduces the work per sample and
allows more samples to run concurrently.
Our scaling model predicts when this additional benefit reverses the preferred unraveling between execution
limited by workers and execution limited by memory.
Benchmarks across several noise channels illustrate how physical and numerical parameters change the bond
dimension and sampling requirements.
As an application, a pilot protocol estimates the inputs to this regime classification for a specified
problem.
These results show how the computational cost of an unraveling depends jointly on trajectory complexity,
statistical convergence, and available hardware resources.
\end{abstract}

\maketitle

\section{Introduction}
The accurate simulation of open quantum systems constitutes a central challenge in quantum science,
underpinning the modeling and validation of noisy quantum devices, many-body dynamics, and near-term quantum
algorithms.
This perspective is grounded in the fundamental observation that every quantum system is, to some extent, open
and inevitably interacts with its environment.
In the context of quantum computing, noise represents a principal obstacle.
It is therefore crucial not only to understand in detail how quantum noise affects coherent dynamics, but also
to delineate the regimes of quantum evolution that can be effectively ``dequantized'' through
advanced classical simulation techniques \cite{Zhou_2020}.
Over the past decade, applying advanced classical simulation methods to quantum simulation platforms of
steadily increasing capability has proven to be a highly fruitful endeavor
\cite{Zhou_2020,PRXQuantum.5.010308,Trotzky}: This helps assessing the computational power of quantum
simulators and to turn paradigmatic settings into quantum technological devices \cite{MindTheGaps}.
In such efforts, tensor network methods take center stage.
Indeed, in many-body settings, tensor network methods---most prominently \emph{matrix product state} (MPS)
techniques for one spatial dimension---provide a scalable classical
framework by exploiting the typically limited entanglement structure of
relevant quantum states~\cite{White1992, vidal_2004, Schollwock_2011,AreaReview,
RevModPhys.93.045003,Orus-AnnPhys-2014}.
For closed systems, the computational cost of MPS simulations is closely tied to entanglement growth, as
captured by the bond dimension.

Among possible model descriptions, Markovian or Lindbladian open quantum systems are particularly important.
They capture forgetful quantum dynamics, which is ubiquitous for weakly
coupled quantum systems and appears naturally in many physical settings.
Open system dynamics governed by Lindblad master equations can be simulated using trajectory-based approaches,
which unravel the mixed state evolution into ensembles of stochastic pure state
trajectories~\cite{Dalibard_1992, Molmer_92, Carmichael_1993, Wiseman_Milburn_2009}.
In MPS-based implementations, each stochastic trajectory is represented as an individual many-body pure state
whose time evolution is sampled repeatedly to reconstruct physical observables.
The computational object of interest is therefore not the density matrix itself, but the ensemble of
stochastic trajectories that approximate it.
Consequently, the relevant notion of entanglement for assessing trajectory-level simulation complexity is that
of the individual trajectories rather than that of the mixed state generated by the Lindblad dynamics.

A substantial body of work has investigated how specific stochastic unravelings and measurement-induced
dynamics influence entanglement along individual trajectories.
In particular, continuous monitoring schemes and suitably chosen unravelings can significantly suppress
trajectory entanglement or alter its scaling
behavior~\cite{block_2022, liu_2024, cecile_2024, guimaraes_2023, leung_2024, kshetrimayum_2017}.
More recently, Vovk and Pichler have systematically explored the freedom inherent in stochastic unravelings,
introducing
entanglement-optimized and conventional schemes and demonstrating that unraveling choice can qualitatively
reshape trajectory entanglement scaling, including finite-size transitions between area- and volume-law
behavior and exponential cost separations between MPS trajectories and \emph{matrix product operator}
(MPO)-based simulations in specific models~\cite{vovk_2022, vovk_2024}.
A rigorous underpinning of entanglement-optimal mixed state decompositions has been delivered in Ref.\
\cite{SimonSimulation}.
These results and rigorous insights establish that unraveling choice can dramatically modify trajectory
entanglement properties, suggesting that apparent simulation cost may depend sensitively on how the master
equation is decomposed.

The cost of a trajectory simulation includes storage, time evolution, and statistical sampling.
These contributions have been discussed in prior
work~\cite{daley_2014, transchel_2014, weimer_2021, moroder_2023,SimonSimulation}.
The bond dimension sets the memory required by each MPS trajectory, while bond dimension and numerical
resolution determine the work needed to evolve it.
Sampling variance determines how many trajectories are needed for a specified observable accuracy.
Finite memory can also restrict parallel execution even when additional workers are available.
Changing the unraveling can therefore change both the work per trajectory and the number of trajectories that
can run concurrently.
The question is how this coupling affects the relative computational cost of equivalent unravelings at fixed
physical dynamics and accuracy.

In this work, we characterize computational regimes in MPS-based quantum trajectory simulations.
Rather than seeking a universal ranking of unravelings, we ask how physical noise parameters, numerical
choices, and hardware constraints redistribute computational effort at fixed accuracy.
We describe relative memory, runtime, and sampling requirements using two dimensionless ratios of trajectory
bond dimensions and required sample counts.
These ratios define a decision geometry for comparing physically equivalent unravelings under finite memory
and parallelism.
As an application, we describe a pilot protocol for estimating these inputs for a specified simulation.

To reach system sizes beyond exact state-vector methods, we employ the \emph{tensor jump method} (TJM), a
trajectory formulation of Lindblad dynamics that evolves ensembles of MPS pure states and avoids the
exponential overhead of density-matrix simulation~\cite{sander_TJM_2025}.
Here, TJM serves as a scalable benchmarking platform for studying how trajectory bond dimension, stochastic
sampling, and finite computational resources jointly determine simulation
cost~\cite{daley_2014, transchel_2014, weimer_2021, moroder_2023, zwolak_2004}.
We benchmark one-dimensional spin models with local depolarizing noise up to 80 sites and complement these
results with Heisenberg-model case studies under additional noise channels.

Our cost model predicts that two equivalent unravelings can exchange their relative computational advantage
when execution changes from being limited by workers to being limited by memory, even for the same physical
dynamics and accuracy requirement.
The numerical benchmarks resolve the separate contributions from bond dimension and sampling.
They show that reduced bond dimension can be accompanied by increased sampling requirements or finer time step
resolution.
The hardware projections illustrate the resulting resource dependence within the scaling model.

\section{Quantum trajectory simulations}
This section introduces the theoretical and algorithmic framework underlying quantum trajectory simulations of
open quantum systems.
We summarize the Lindblad formalism, its stochastic unravelings, and the numerical parameters that control
trajectory evolution.
No assumptions about computational cost or scaling behavior are being made at this stage.

\subsection{Lindblad dynamics and quantum trajectories}
The interaction of a quantum system with a weakly coupled, memoryless environment is commonly modeled by a
Lindblad master equation~\cite{Lindblad1976, Breuer_2007}.
Let $\rho \in \mathcal{B}(\mathcal{H})$ denote the system density operator and
$H_0 \in \mathcal{B}(\mathcal{H})$ the closed system Hamiltonian acting on a Hilbert space $\mathcal{H}$.
The Lindblad equation reads
\begin{equation}
\label{eq:Lindblad}
\frac{d}{dt}\rho
= -i[H_0,\rho]
+ \sum_{m=1}^k \gamma_m
\left(
L_m \rho L_m^\dagger
- \frac{1}{2}\{L_m^\dagger L_m, \rho\}
\right),
\end{equation}
where $\{L_m\}_{m=1}^k \subset \mathcal{B}(\mathcal{H})$ are jump operators describing dissipative processes
such as relaxation, excitation, or dephasing, and $\{\gamma_m\}_{m=1}^k$ are the corresponding non-negative
dissipation rates.

The direct integration of Eq.~\eqref{eq:Lindblad} is, however, generally computationally intractable
already for moderately sized quantum systems, i.e., it scales exponentially with system size due to the
dimension of $\rho \in \mathbb{C}^{d^L \times d^L}$ where $d$ is the physical dimension and $L$ is the system
size.
Trajectory-based approaches avoid this bottleneck by unraveling the Lindblad dynamics into ensembles of
stochastic pure state trajectories~\cite{Dalibard_1992, Carmichael_1993, Molmer_92, Wiseman_Milburn_2009}.
Physical observables are recovered by averaging over trajectories.

In the specific \emph{Monte Carlo wavefunction} (MCWF) formalism, each trajectory evolves under an effective
non-Hermitian Hamiltonian
\begin{equation}
\label{eq:Heff}
H = H_0 - \frac{i}{2}\sum_{m=1}^k \gamma_m L_m^\dagger L_m,
\end{equation}
interrupted by stochastic quantum jumps.
Over a time step $\delta t$, the
state vector evolves according to
\begin{equation}
\ket{\Psi(t+\delta t)} = e^{-iH\delta t}\ket{\Psi(t)},
\end{equation}
with a total jump probability
\begin{equation}
\delta p(t)
= \sum_{m=1}^k \delta t\, \gamma_m
\langle \Psi(t) | L_m^\dagger L_m | \Psi(t) \rangle.
\end{equation}
If a jump occurs, a jump operator $L_m$ is applied with probability proportional to its contribution to
$\delta p(t)$, otherwise, the state vector is renormalized.
Repeating this procedure generates a stochastic trajectory $\ket{\Psi_j(t)}$.
Ensemble
averages approximate expectation values,
\begin{equation}
\langle O(t) \rangle
= \frac{1}{N}\sum_{j=1}^N
\bra{\Psi_j(t)} O \ket{\Psi_j(t)} ,
\end{equation}
with statistical convergence governed by standard Monte Carlo sampling, i.e., $\mathcal{O}(1/\sqrt{N})$ with a
variance-dependent pre-factor, as a faithful estimator for the mixed quantum state at time $t$.

\subsection{Unraveling freedom}
The stochastic representation of Lindblad dynamics is not unique.
A given Lindblad equation admits infinitely many stochastic unravelings, corresponding to different ways of
resolving environmental interactions at the level of individual trajectories~\cite{Wiseman_Milburn_2009}.
While all valid unravelings reproduce the same mixed state dynamics upon ensemble averaging, they generally
generate distinct trajectory ensembles.

From a physical perspective, different unravelings are commonly associated with different measurement or
monitoring schemes applied to the environment, such as quantum jump, diffusive, or hybrid
descriptions~\cite{Wiseman_Milburn_2009, daley_2014}.
This freedom has been extensively explored in the context of measurement-induced dynamics and
monitoring-driven entanglement phenomena, where unraveling choice strongly affects trajectory-level
properties~\cite{block_2022, liu_2024, guimaraes_2023, leung_2024}.

The freedom in choosing stochastic unravelings reflects the non-uniqueness of the Lindblad operator
decomposition, i.e., different sets of jump operators and associated monitoring schemes generate distinct
ensembles of pure state trajectories while reproducing the same mixed state evolution.
Vovk and
Pichler have systematically explored this freedom, demonstrating that appropriately chosen unravelings can
alter trajectory entanglement scaling and even induce finite-size transitions between area- and volume-law
behavior~\cite{vovk_2022, vovk_2024,AreaReview}.
Here, rather than seeking entanglement-minimizing unravelings, we treat this as a computational degree of
freedom and analyze how different choices redistribute cost across bond dimension, runtime, and sampling
requirements.

Although different unravelings reproduce the same Lindblad dynamics at the ensemble level, in the sense that
they all give rise to faithful estimators of the quantum state at a given time, they are implemented in
practice through concrete choices of dissipation rates, time step discretization, and stochastic update rules.
A clear distinction between physical parameters and numerical control parameters is therefore crucial before
their computational implications can be meaningfully addressed.

\subsection{Physical and numerical parameters}
Quantum trajectory simulations are governed by both physical and numerical parameters that control how the
continuous-time Lindblad dynamics is realized algorithmically.
The dissipation rates $\{\gamma_m\}_{m=1}^k$ appearing in Eq.~\eqref{eq:Lindblad} are physical parameters
fixed by the model.
They determine the strength and relative weight of different noise channels and fully specify the Lindblad
dynamics~\cite{Lindblad1976, Breuer_2007}.
The time step $\delta t$ is a numerical parameter that controls the temporal discretization of the stochastic
evolution.
In trajectory-based simulations, $\delta t$ sets both the resolution of the numerical integrator and the
frequency with which the system is sampled for quantum jumps~\cite{Wiseman_Milburn_2009, daley_2014}.
For a convergent scheme with other numerical errors controlled, time step refinement approaches the same
Lindblad dynamics in expectation, while individual stochastic realizations can differ.
The effective jump probability
\begin{equation}
\delta p \propto \gamma\, \delta t
\end{equation}
provides an emergent algorithmic measure of stochasticity per time step, up to state-dependent expectation
values of the jump operators.
Unlike $\gamma$, which is a physical parameter, or $\delta t$, which is numerical, $\delta p$ is not an
independent control parameter but reflects their combined effect.
In practice, $\delta p$ is frequently used as a heuristic proxy for noise strength in trajectory simulations.
However, as we analyze explicitly in this work, $\delta p$ does not uniquely characterize trajectory behavior
across different discretizations, even when ensemble dynamics coincide.

\subsection{Tensor network representation of trajectories}
Trajectory methods replace the exponential complexity of density matrix evolution by an approximate ensemble
of pure states, reducing the memory scaling from $\mathcal{O}(d^{2L})$ to $\mathcal{O}(N d^L)$.
For large systems, this scaling remains prohibitive and motivates the use of tensor network representations
for individual trajectories.
In one spatial dimension, state vectors $\ket{\Psi} \in \mathbb{C}^{d^L}$ can be efficiently represented as
\emph{matrix product state} (MPS) vectors~\cite{White1992, Vidal_2003, Schollwock_2011},
\begin{equation}
\ket{\Psi}
= \sum_{\sigma_1,\ldots,\sigma_L}
M^{\sigma_1}_1 \cdots M^{\sigma_L}_L
\ket{\sigma_1,\ldots,\sigma_L},
\end{equation}
where $M^{\sigma_\ell}_\ell \in \mathbb{C}^{\chi_{\ell-1} \times \chi_\ell}$ are local tensors and
$\{\chi_\ell\}$ are bond dimensions encoding bipartite entanglement with boundaries $\chi_0=\chi_L=1$.
When entanglement remains moderate, this representation yields polynomial scaling in system size with memory
cost $\mathcal{O}(d L \chi^2)$ for a single trajectory.
The computational cost of evolving an MPS typically scales with a higher power of the bond dimension, commonly
as $\mathcal{O}(\chi^3)$ per local update, reflecting the cost of tensor contractions and local linear
algebra, with the precise pre-factor depending on the chosen time-evolution scheme
\cite{Schollwock_2011, Paeckel_2019}.

At fixed system size, a full vector MCWF trajectory stores $d^L$ amplitudes irrespective of its unraveling.
Such trajectories can already run in parallel, with concurrency limited by memory as well as available
workers.
Unraveling choice can still affect runtime and sampling variance, but does not change this leading
state storage requirement.
On the other hand, for MPS trajectories, storage depends on the unraveling through $\chi^2$.
Changing the unraveling can therefore change both the work per trajectory and the number of trajectories
that fit concurrently in memory.

In this work, we employ the \emph{tensor jump method} (TJM)~\cite{sander_TJM_2025}, which combines stochastic
quantum trajectories with matrix product state representations and \emph{time-dependent variational principle}
(TDVP) evolution~\cite{Haegeman2011, Haegeman_2016, Paeckel_2019}.
Between stochastic jump events, the state is propagated within the MPS manifold using TDVP, while quantum
jumps are implemented as local operator insertions followed by canonicalization and controlled truncation.
This structure enables large ensembles of trajectory simulations while maintaining a transparent separation
between physical dynamics and numerical approximation.

The framework of quantum trajectories together with an MPS representation introduces a concrete set of
resource controls.
The bond dimension $\chi$ governs the memory footprint of individual trajectories and the cost of local tensor
network updates, while the number of trajectories $N$ governs statistical convergence.
In addition, the time step $\delta t$ controls how many updates are required to reach a fixed physical
evolution time.
These ingredients allow us to formalize the end-to-end computational cost of trajectory simulations in terms
of distinct cost channels.
With the simulation ingredients fixed, we now turn from representation to resource accounting.

\section{Computational cost structure}
\label{sec:cost_structure}
Given an MPS representation of trajectories, the computational cost of a fixed-accuracy simulation is not
governed by a single parameter.
Instead, it separates into three distinct contributions: per-trajectory memory cost, per-trajectory runtime
cost, and sampling cost, the latter reducible through parallel execution of independent trajectories.

To compare physically equivalent stochastic unravelings within this multi-channel structure, we introduce two
dimensionless ratios, a bond dimension inflation factor $\alpha$, capturing relative per-trajectory memory and
runtime scaling, and a sampling inflation factor $\kappa$, quantifying relative sampling effort at fixed
accuracy.
We then derive a hardware-aware criterion that determines which unraveling is favored under given memory and
parallelism constraints.

In practice, MPS-based trajectory simulations approximate the exact Lindblad dynamics through a combination of
finite time step integration, projection onto a finite-bond dimension variational manifold, and finite
stochastic sampling.
Throughout this work, ``fixed accuracy'' means convergence with respect to the numerical controls relevant for
the comparison, e.g., the time step $\delta t$, the MPS truncation criterion, and the statistical uncertainty
of the trajectory estimator.
For large systems, where exact reference evolution is unavailable, the sampling component refers to the
estimated standard error of the trajectory average rather than a directly measured absolute error to the exact
Lindblad solution.

\subsection{Three canonical cost channels}
We consider trajectory-based simulations in which expectation values are estimated from an ensemble of $N$
independent trajectories, each represented as an MPS with characteristic maximum bond dimension $\chi$.
The total cost then consists of the following components:
\begin{enumerate}
\item {\textbf{Per-trajectory memory cost (MPS storage).}}
For a one-dimensional MPS, the memory footprint of a single trajectory scales as
\begin{equation}
\mathrm{mem}(\chi) \propto L\, d\, \chi^2,
\label{eq:mem_scaling_def}
\end{equation}
up to implementation-dependent constants~\cite{Schollwock_2011}.

\item{\textbf{Per-trajectory runtime cost (MPS time evolution).}}
For common MPS time evolution schemes based on local tensor contractions and factorizations, the dominant cost
per local update scales cubically with bond dimension~\cite{Paeckel_2019, Schollwock_2011}, primarily due to
local tensor contractions and factorizations such as singular value decompositions.
We model the per-trajectory runtime to reach a fixed physical time $T$ as
\begin{equation}
\mathrm{time}(\chi)
\propto
\frac{T}{\delta t}\, L\, d\, \chi^3,
\label{eq:time_scaling_def}
\end{equation}
where the factor $T/\delta t$ counts the number of time steps.

\item{\textbf{Sampling cost (number of trajectories).}}
Observable estimates converge with the number of trajectories at standard Monte Carlo rates, with a pre-factor
that depends on the variance of the trajectory ensemble.
For an observable $O$ and unraveling $U$, the standard error of the trajectory estimator is
\begin{equation}
\Delta O_U
\approx
\sqrt{\frac{\mathrm{Var}_U[O]}{N}},
\end{equation}
where $\mathrm{Var}_U[O]$ is the unraveling-dependent trajectory-to-trajectory variance.
For a prescribed target statistical uncertainty $\varepsilon$, the required number of trajectories is denoted
$N_U(\varepsilon)$.
Different stochastic unravelings generate distinct trajectory ensembles and therefore generally exhibit
different sampling variances, directly affecting the trajectory count required to reach fixed statistical
accuracy.
\end{enumerate}

These three channels are conceptually distinct.
Memory and runtime are controlled primarily by the bond dimension of \emph{individual} trajectories, whereas
sampling cost is controlled by trajectory-to-trajectory fluctuations of the \emph{ensemble}.
While absolute pre-factors depend on implementation details, such as local update kernels, the number of jump
operators, and linear-algebra backends, the separation into memory, runtime, and sampling costs is the
resource-accounting structure used throughout this work.

\subsection{Inflation ratios and the $(\alpha,\kappa)$ characterization}
We compare two physically equivalent stochastic unravelings, denoted $A$ and $B$, simulated to the same target
accuracy $\varepsilon$ for identical Lindblad dynamics.
In the ideal limit of infinite sampling, converged time step, and controlled MPS truncation, both unravelings
reproduce the same Lindblad expectation values.
At finite computational resources, however, they may differ in trajectory bond dimension, estimator variance,
and therefore total cost.

Let $\chi_A$ and $\chi_B$ denote the maximum bond dimensions reached by the two unravelings, and let
$N_A(\varepsilon)$ and $N_B(\varepsilon)$ be the numbers of trajectories required to reach the same target
statistical uncertainty $\varepsilon$ for the chosen observable.
We define two dimensionless ratios
\begin{equation}
\alpha \coloneqq \frac{\chi_A}{\chi_B},
\qquad
\kappa(\varepsilon) \coloneqq  \frac{N_B(\varepsilon)}{N_A(\varepsilon)},
\label{eq:alpha_kappa_def}
\end{equation}
which quantify how computational effort is redistributed between the two unravelings.

The ratio $\alpha$ measures relative per-trajectory complexity for the specified pair of unravelings and
simulation scenario.
Since MPS memory scales as $\chi^2$ and dominant runtime costs scale approximately as $\chi^3$, $\alpha$
determines the relative memory footprint and per-trajectory runtime overhead within this cost model.
The ratio $\kappa$ measures the relative sampling cost required to reach fixed statistical accuracy, encoding
differences in trajectory-to-trajectory variance.
Neither ratio is an intrinsic property of a noise model or an unraveling alone; both are relative,
scenario-dependent quantities.

The relative ordering of $\alpha$ and $\kappa$ leads to four qualitatively distinct regimes.
When $\alpha$ and $\kappa$ lie on opposite sides of unity, one unraveling strictly dominates the other across
both bond dimension and sampling requirements.
For example, $\alpha>1$ and $\kappa<1$ implies that unraveling $B$ has both lower per-trajectory complexity
and lower sampling cost, and is therefore preferred independently of hardware constraints; the case $\alpha<1$
and $\kappa>1$ analogously favors unraveling $A$.
In contrast, when $\alpha$ and $\kappa$ lie on the same side of unity, the unravelings trade per-trajectory
complexity against sampling effort, and the preferred strategy becomes hardware-dependent.

Throughout this work, we primarily examine the trade-off regime in which unraveling~$A$ yields larger bond
dimensions but faster statistical convergence, while unraveling~$B$ produces less entangled trajectories at
the expense of increased sampling effort.
Under this ordering, $\alpha \ge 1$ and $\kappa \ge 1$, and neither unraveling trivially dominates all cost
channels.
The framework introduced here therefore classifies unravelings not by a universal ranking, but by their
location within this two-parameter cost plane.

While the analytical decision geometry in the $(\alpha,\kappa)$ plane is fixed, the numerical values of
$\alpha$ and $\kappa$ associated with a given pair of unravelings are not intrinsic constants.
Rather, they depend on physical parameters and numerical choices such as the Hamiltonian, noise model,
observable, target accuracy, time step discretization, and system size.
The role of the following sections is therefore to determine how concrete simulation scenarios populate the
$(\alpha,\kappa)$ plane and to identify the parameter regimes in which unraveling choice becomes
hardware-dependent.
\begin{figure}[t!]
    \centering
    \includegraphics[width=\linewidth]{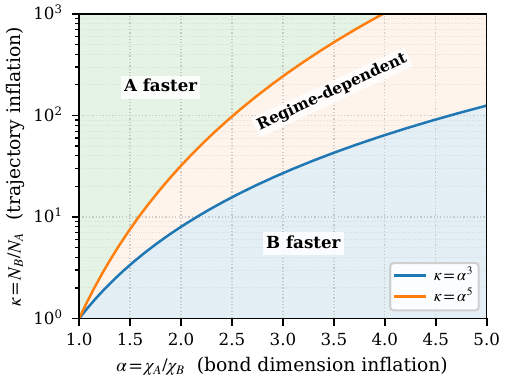}
    \caption{
    \textbf{Analytical redistribution of computational cost between bond dimension and sampling requirements.}
    Decision geometry in the $(\alpha,\kappa)$ plane for two physically equivalent unravelings $A$ and $B$,
    where $\alpha=\chi_A/\chi_B$ quantifies relative per-trajectory complexity and $\kappa=N_B/N_A$ quantifies
    relative sampling cost.
    The curves $\kappa=\alpha^3$ and $\kappa=\alpha^5$ correspond to the decision boundaries in the
    thread-limited and memory-limited regimes, respectively.
    Below each curve, the reduction in per-trajectory cost associated with smaller bond dimension outweighs
    the increased number of trajectories; above it, the sampling overhead dominates.
    For $\alpha>1$, the preferred unraveling differs between these two limiting regimes when
    $\alpha^3<\kappa<\alpha^5$.
    }
    \label{fig:ak_plot}
\end{figure}

\subsection{Decision boundaries under realistic hardware constraints}
Quantum trajectory simulations are naturally amenable to parallel execution, since individual trajectories are
statistically independent and can be evolved concurrently.
To translate the abstract cost ratios $(\alpha,\kappa)$ into an end-to-end runtime comparison, we model
trajectory execution on hardware with a finite parallelism budget of $p$ workers and a total memory budget
$M$.
These constraints determine how many of the trajectories required for a target accuracy can be evolved
simultaneously and, in large-ensemble settings, whether execution is primarily limited by worker count or by
memory capacity.

We compare the same physical dynamics and observable at fixed system size, evolution time, time step, and
target accuracy.
The model maintains the dominant quadratic dependence of trajectory storage and cubic dependence of
trajectory runtime on bond dimension, while omitting implementation-dependent prefactors.

Let $N_j(\varepsilon)$ denote the number of trajectories required by unraveling $j$ to reach the target
statistical accuracy $\varepsilon$.
This is not a hardware resource, but a property of the stochastic estimator for the chosen observable and
accuracy target.
Let $m_j$ denote the maximum number of trajectories of unraveling $j$ that can be stored concurrently in
memory, with $m_j$ scaling as
\begin{equation}
m_j \propto \frac{M}{L\, d\, \chi_j^2}.
\label{eq:mj_def}
\end{equation}
The effective parallelism for unraveling $j$ is then limited by the minimum of available workers, required
trajectories, and memory capacity,
\begin{equation}
P_j = \min\!\bigl(p,\; N_j(\varepsilon),\; m_j\bigr).
\label{eq:Pj_def_cost}
\end{equation}
Using the scaling in Eq.~\eqref{eq:time_scaling_def}, the total wall time can be modeled as
\begin{equation}
T_j \propto \frac{N_j(\varepsilon)}{P_j}\,\frac{T}{\delta t}\,L\,d\,\chi_j^3,
\label{eq:Tj_cost_model}
\end{equation}
which accounts for batched concurrent execution of independent trajectories.
This leads to the runtime ratio for a fixed system size and time step
\begin{equation}
\frac{T_A}{T_B}
\propto
\frac{\alpha^3}{\kappa}\,\frac{P_B}{P_A}.
\label{eq:TA_TB_cost}
\end{equation}

Equation~\eqref{eq:Pj_def_cost} also includes the fully concurrent regime in which all required trajectories
fit within both the worker and memory budgets.
If $N_j(\varepsilon)\le p$ and $N_j(\varepsilon)\le m_j$, then $P_j=N_j(\varepsilon)$ and all trajectories for
unraveling $j$ can be executed in a single batch.
In this case the required trajectory count still determines total computational work, but it no longer
increases the idealized wall time.
If both unravelings are fully concurrent, the wall-time comparison reduces to the per-trajectory cost ratio,
$T_A/T_B\propto\alpha^3$.
The hardware-dependent comparison between worker count and memory capacity becomes relevant once
$N_j(\varepsilon)$ is large enough that at least one of the concurrency constraints, $p$ or $m_j$, is
saturated.
We therefore focus on the two large-ensemble limiting regimes that yield transparent decision boundaries.

\begin{enumerate}
\item \textbf{Thread-limited regime.}
If available workers are the bottleneck ($p \le m_A,m_B$ and $p \le N_A,N_B$), then both unravelings saturate
the same parallelism, $P_A=P_B=p$, and
\begin{equation}
\frac{T_A}{T_B} \propto \frac{\alpha^3}{\kappa}.
\end{equation}
In this regime, trajectories are executed concurrently up to the worker limit, so differences in memory
footprint do not affect the number of simultaneous trajectories.
The competition is therefore between evolving fewer high-cost trajectories and more low-cost trajectories.
Unraveling~$B$ is faster when the increase in trajectory count does not outweigh the reduction in
per-trajectory cost ($\kappa < \alpha^3$), whereas unraveling~$A$ is favored when the additional sampling
overhead dominates ($\kappa > \alpha^3$).

\item \textbf{Memory-limited regime.}
If memory limits concurrency ($m_A,m_B \le p$ and $m_A,m_B \le N_A,N_B$), then the number of trajectories that
can be evolved in parallel is set by memory, $P_j=m_j$.
Using $m_A/m_B \approx \alpha^{-2}$ yields
\begin{equation}
\frac{T_A}{T_B} \propto \frac{\alpha^5}{\kappa}.
\end{equation}
Here, the smaller bond dimension of unraveling~$B$ allows significantly more trajectories to fit in memory and
be processed concurrently.
In this regime, having more required trajectories is partially compensated by increased concurrency, while the
larger bond dimension of unraveling~$A$ restricts the number of simultaneous trajectories.
As a result, unraveling~$B$ is favored whenever $\kappa < \alpha^5$.
\end{enumerate}

For large ensembles, $T_A/T_B\propto\alpha^3/\kappa$ when both unravelings saturate the worker limit, while
when both saturate the memory limit, $T_A/T_B\propto\alpha^5/\kappa$.
The additional factor $\alpha^2$ in the latter limit comes from the increased concurrency permitted by the
smaller trajectory representation, and the curves $\kappa=\alpha^3$ and $\kappa=\alpha^5$ therefore define the respective decision boundaries.
For $\alpha>1$ and $\alpha^3<\kappa<\alpha^5$, unraveling~$A$ is favored in the worker-limited regime and
unraveling~$B$ in the memory-limited regime, while outside this interval, the same unraveling is favored in both saturated limits.
Mixed and fully concurrent cases remain described by Eq.~\eqref{eq:TA_TB_cost}.

\begin{figure*}
    \centering
    \includegraphics[width=\linewidth]{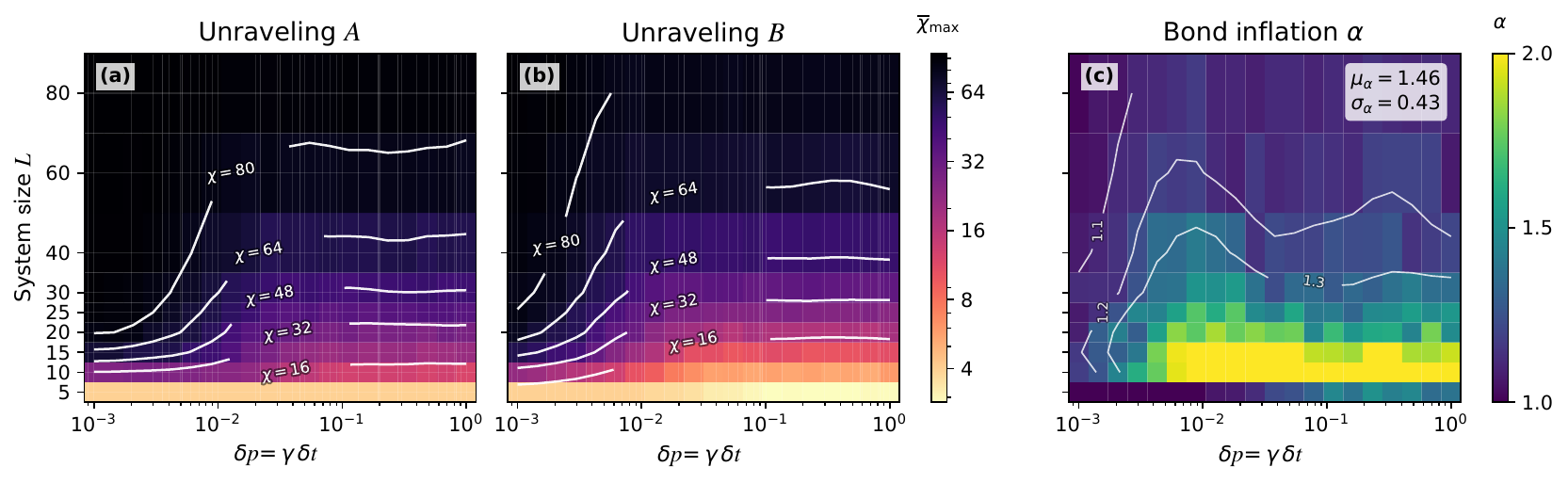}
    \caption{
    \textbf{Finite-size bond dimension trends in the TJM depolarizing benchmark.}
    \textbf{(a,b)} Peak average trajectory bond dimension $\overline{\chi}$ as a function of system size $L$
    and discretized noise parameter $\delta p=\gamma \delta t$ for (a) Pauli-jump and (b) measurement-based
    unravelings.
    White contours denote constant $\chi_{\mathrm{max}}$.
    \textbf{(c)} Bond inflation ratio $\alpha = \chi_{\mathrm{max},A}/\chi_{\mathrm{max},B}$ as a function of
    system size $L$ and $\delta p$.
    Results are obtained for a transverse-field Ising chain initialized in a N\'eel state, evolved to $T=5$
    with $\delta t=0.1$ under local depolarizing noise using the TJM update rule.
    }
    \label{fig:Large_Scale}
\end{figure*}

The decision geometry visualized in Fig.~\ref{fig:ak_plot} is analytical and independent of any particular
physical model or implementation-specific pre-factors.
What varies between simulation scenarios is therefore not the geometry itself, but the location of a given
problem within the $(\alpha,\kappa)$ plane.
Most importantly, the numerical values of $\alpha$ and $\kappa$ are \emph{not} intrinsic properties of an
unraveling.
They depend on physical parameters such as the Hamiltonian and noise model, as well as numerical
discretization choices and system size.

A central contribution of this work is to demonstrate this regime dependence explicitly.
Using large-scale benchmarks, we show that the same pair of physically equivalent unravelings can occupy
qualitatively different locations in the $(\alpha,\kappa)$ plane as noise strength, time step, and system size
are varied.
As a result, the preferred unraveling can change across parameter regimes even though the underlying decision
geometry remains fixed.
In the following sections, we extract $(\alpha,\kappa)$ for representative unravelings across a range of
physical and numerical regimes.
This allows us to place concrete simulation scenarios within the $(\alpha,\kappa)$ plane and to identify when
apparent cost reductions in one channel are offset by increases in another.

\section{Benchmarking cost redistribution}
\label{sec:benchmarking}

\subsection{Setup}

We now turn to a concrete case study in which we extract the bond dimension and sampling inflation factors
$(\alpha,\kappa)$ for representative stochastic unravelings under controlled physical and numerical
conditions.
The goal of this section is not to identify model-specific phenomena, but to empirically demonstrate how
$\alpha$ and $\kappa$ vary with algorithmic and physical parameters, and how this variation reshapes the cost
trade-offs identified in Sec.~\ref{sec:cost_structure}.

The main benchmarks in this section are performed on the one-dimensional transverse-field Ising model with
open boundary conditions
\begin{equation}
H = -J \sum_{i=1}^{L-1} Z_i Z_{i+1} \;-\; g \sum_{i=1}^{L} X_i ,
\end{equation}
with parameters fixed to the critical point $J=g=1$.
Simulations are initialized in the N\'eel product state vector
\begin{equation}
\ket{\psi_0} = \ket{0,1,0,1\cdots},
\end{equation}
which rapidly generates entanglement under coherent dynamics while avoiding symmetries that could artificially
simplify the evolution.
While we focus on this specific model, the cost-redistribution mechanisms identified below arise from
trajectory-level bond growth, sampling variance, and discretization choices rather than from the detailed
physics of this model alone.

For the main transverse-field Ising benchmarks, we consider open-system dynamics generated by local
depolarizing noise implemented via local jump operators.
To isolate the algorithmic consequences of different stochastic representations of the same Lindblad
generator, we compare two physically equivalent unravelings:
\begin{enumerate}
    \item \textbf{Pauli-jump unraveling (A).}
    Each site undergoes stochastic quantum jumps generated directly by the Pauli operators $X$, $Y$, and $Z$.
    This corresponds to the standard depolarizing channel and produces relatively infrequent but strong state
    updates.

    \item \textbf{Measurement-based unraveling (B).}
    The same depolarizing channel is decomposed into projective measurement outcomes, e.g.,
    $\{\ket{0}\!\bra{0}, \ket{1}\!\bra{1}\}$ for $Z$, $\{\ket{+}\!\bra{+}, \ket{-}\!\bra{-}\}$ for $X$, and
    $\{\ket{+_i}\!\bra{+_i}, \ket{-_i}\!\bra{-_i}\}$ for $Y$.
    This yields more frequent but weaker stochastic updates, modifying individual trajectory dynamics while
    reproducing the same Lindblad evolution upon ensemble averaging.
    To ensure equivalence of the underlying master equation under this decomposition, the corresponding rates
    satisfy $\gamma_B = 2\gamma_A=:2\gamma$.
\end{enumerate}

These two unravelings are not intended to be exhaustive.
Other stochastic representations of Lindblad dynamics have been explored in prior work and can give rise to
qualitatively different trajectory statistics and convergence
behavior~\cite{vovk_2022, vovk_2024, SimonSimulation, AnalogSimQuantinuum}.
Our objective here is not to rank all possible unravelings, but to use a minimal pair to expose mechanisms by
which algorithmic choices redistribute computational cost in MPS-based trajectory simulations.
Unless stated otherwise, the benchmarks in this section use local depolarizing noise.
Other noise models are considered later as additional case studies and may shift the extracted
$(\alpha,\kappa)$ values while leaving the analytical decision geometry unchanged.
Finally, this comparison can be generalized to more than two unravelings.

All simulations are performed using the \emph{tensor jump method} (TJM) \cite{sander_TJM_2025}.
No explicit maximum bond dimension is imposed, so unitary evolution is performed using two-site TDVP without a
hard cutoff on $\chi$.
Instead, bond growth is controlled through a discarded-weight truncation criterion, i.e., after each two-site
update, singular values are truncated using a threshold of $s_\text{max}=10^{-6}$.
The bond dimension therefore adapts dynamically in response to both coherent evolution and stochastic jumps.

Expectation values are obtained by averaging over stochastic trajectories.
In the ideal limit of infinite trajectory number, converged time step, and controlled MPS truncation, both
valid unravelings reproduce the same Lindblad expectation values.
At finite numerical resources, however, the two unravelings can differ in trajectory bond dimension, estimator
variance, and therefore total computational cost.

Throughout this work, fixed accuracy means convergence with respect to the numerical controls relevant for the
comparison.
We distinguish three contributions.
First, time step bias must be checked by decreasing $\delta t$ for the physical parameters and observables
being compared.
Second, the discarded weight threshold controls MPS truncation, but does not by itself bound the total MPS
approximation error such that we assume this error is subdominant for the reported comparisons.
Third, sampling error is estimated from trajectory-to-trajectory fluctuations of the measured observables.
For an observable $O$ and unraveling $U$, this gives a standard error of the mean of
\begin{equation}
\Delta O_U \approx \frac{\sigma_U[O]}{\sqrt{N}},
\end{equation}
where $\sigma_U[O]$ is the empirical standard deviation across trajectories.
For large systems, where exact reference evolution is unavailable, this statistical error is an estimator
uncertainty rather than a directly measured absolute error to the exact Lindblad solution.
In the small-system convergence benchmarks below, exact reference evolution is available, and we additionally
use the absolute deviation from the exact result to extract trajectory counts at a prescribed accuracy.

All benchmarks are performed using \textsc{MQT-YAQS}~\cite{YAQS, wille_mqt2024}, our MPS-based framework for
stochastic and time-dependent quantum simulation.
Results are obtained on a desktop equipped with an Intel Core i7-13600KF CPU with 20 threads and 64 GB of RAM.
Within each comparison, identical build settings, linear algebra backends, and threading configurations are
used to ensure that observed differences reflect algorithmic behavior rather than environment-dependent
performance artifacts.
Unless otherwise specified, bond dimensions are reported as the maximum bond dimension reached over the course
of the evolution.

\begin{figure*}
    \centering
  \includegraphics[width=\linewidth]{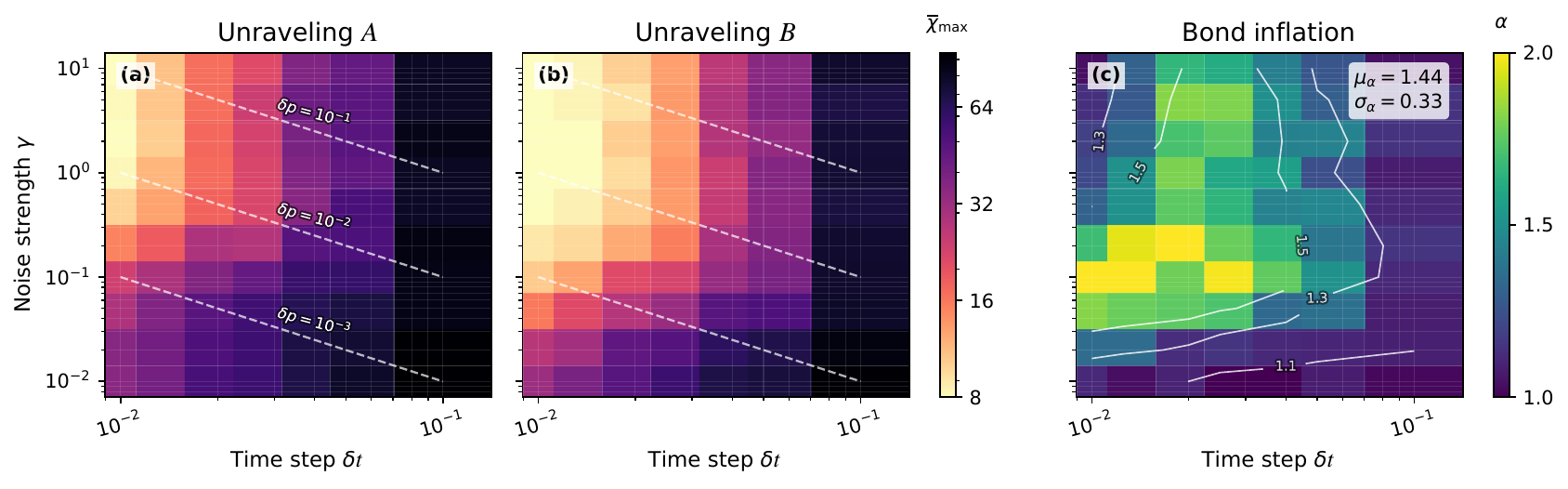}
    \caption{
    \textbf{Monitoring-resolution dependence of bond dimension in $(\gamma,\delta t)$ space.}
    \textbf{(a,b)} Peak trajectory bond dimension $\chi$ for $L=65$ as a function of noise strength $\gamma$
    and time step $\delta t$ for unravelings A and B.
    Dashed lines indicate contours of constant $\delta p=\gamma\delta t$.
    \textbf{(c)} Bond inflation ratio $\alpha=\chi_A/\chi_B$ as a function of $\gamma$ and $\delta t$.
    Results are shown up to $T=5$ using $N=30$ trajectories to characterize systematic bond dimension trends.
    }
    \label{fig:gamma_dt}
\end{figure*}

\begin{figure}
    \centering
  \includegraphics[width=.9\linewidth]{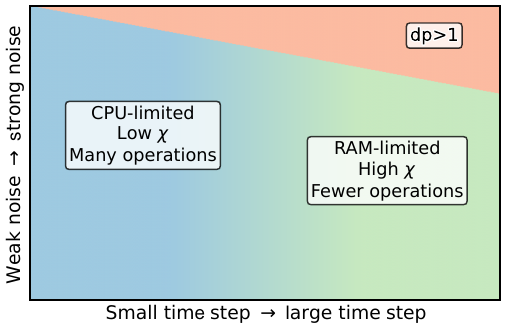}
    \caption{
    \textbf{Schematic discretization-controlled computational regimes.}
    Conceptual map in $(\gamma,\delta t)$ space summarizing the behavior in \cref{fig:gamma_dt}.
    For small $\delta t$, fine monitoring suppresses trajectory bond dimension $\chi$ but increases the number
    of updates required to reach fixed physical time.
    For larger $\delta t$, bond dimension becomes weakly-dependent on $\gamma$ and memory usage dominates
    resource requirements.
    Boundaries are schematic and indicate smooth crossovers rather than sharp transitions.
    }
    \label{fig:gamma_dt_summary}
\end{figure}

\subsection{Finite-size trends of trajectory bond dimension}
A common intuition in open-system simulation is that environmental noise suppresses entanglement and thereby
reduces the bond dimension required to represent quantum states.
In trajectory-based methods, this expectation is often transferred to the entanglement of individual
stochastic trajectories, suggesting that stronger noise may reduce the cost of MPS-based simulations.
Here we test this intuition within the present TJM/MCWF-style benchmark by resolving the trajectory-level
maximum bond dimension $\chi_{\mathrm{max}}$ as a function of system size $L$.
This quantity directly controls the per-trajectory memory footprint and contributes to the per-trajectory
runtime cost.

All results in this subsection are obtained for fixed physical evolution time $T=5$ and time step
$\delta t=0.1$, using $N=30$ stochastic trajectories.
While this trajectory count is insufficient for fully converged observables, it captures systematic
finite-size trends in $\chi$ as a function of $L$ and noise strength.
Increasing $N$ reduces statistical fluctuations in these estimates but does not change the benchmark role of
this subsection, which is to characterize the bond dimension component of the cost model.

\cref{fig:Large_Scale}\textbf{(a,b)} shows the trajectory-averaged peak bond dimension $\overline{\chi}$ as a
function of system size $L$ and effective jump probability $\delta p=\gamma\delta t$ for two physically
equivalent unravelings.
We fix $\delta t$ and vary the physical noise strength $\gamma$, thereby isolating how trajectory complexity
responds to increasing stochasticity at fixed numerical resolution.
Although the continuous-time Lindblad limit formally requires small per-step jump probabilities, we also
include larger values of $\delta p$ as an algorithmic probe of stochastic update strength.
For sufficiently large $\delta p$, the discrete-time dynamics no longer represents a strict continuous-time
unraveling without resolving multiple jumps per time step, and these points should therefore be interpreted
only as a probe of the implemented update rule.

At small and intermediate system sizes, increasing $\delta p$ reduces $\overline{\chi}$ for both unravelings,
consistent with the standard intuition that stochastic jumps interrupt coherent entanglement growth along
individual trajectories.
The more relevant feature of \cref{fig:Large_Scale}, however, is the finite-size behavior of this reduction.
Over the accessible finite-time system-size range, the noise-induced suppression of bond dimension becomes
progressively less pronounced as $L$ increases, and $\overline{\chi}$ approaches a plateau-like regime that
depends only weakly on $\delta p$ for the parameters studied here.

This plateau-like behavior should be interpreted as a benchmark-specific finite-size trend rather than as an
asymptotic statement about arbitrary MPS trajectory methods.
In particular, the behavior reported here is tied to the TJM/MCWF-style update structure, the finite evolution
time, the transverse-field Ising dynamics considered, and the two unravelings compared.
The low-noise dynamics already exhibits limited bond growth over the accessible range, so part of the apparent
saturation may reflect the finite-time entanglement structure of the underlying model rather than a universal
limitation imposed by trajectory sampling.
Other trajectory constructions, including Kraus- or reset-based representations, may therefore produce
qualitatively different finite-size behavior and can in some limits drive trajectories closer to product-state
structure.

The quantitative contrast between unravelings is summarized in \cref{fig:Large_Scale}\textbf{(c)}, which shows
the bond inflation ratio $\alpha = \chi_A/\chi_B$.
Moderate inflation ($\alpha \sim 1.2$--$1.5$) appears at intermediate system sizes and noise strengths,
indicating that unraveling choice can meaningfully affect per-trajectory memory cost in this regime.
As $L$ increases, $\alpha$ narrows toward unity in the present benchmark.
Since $\alpha$ is a relative comparison between the two specified unravelings, this narrowing should not be
read as an absolute statement that unraveling choice becomes irrelevant in all large systems.
Rather, it reflects the plateau-like bond dimension behavior observed for this finite-time benchmark and the
particular TJM update rule used here.

These results show that reduced trajectory bond dimension at small and intermediate system sizes does not
automatically imply a persistent reduction in per-trajectory cost across the accessible system-size range.
More generally, they illustrate why bond dimension alone is not sufficient to assess the total cost of a
trajectory simulation.
Even when noise suppresses trajectory entanglement, the magnitude and persistence of this reduction depend on
the stochastic representation, numerical resolution, and finite-time entanglement structure of the simulated
model.

Finally, although \cref{fig:Large_Scale} parametrizes noise via $\delta p=\gamma\delta t$, the present
analysis isolates only the bond dimension component of computational cost.
Total runtime additionally depends on trajectory sampling requirements, quantified by $\kappa$.
We therefore turn next to extracting the joint $(\alpha,\kappa)$ characterization across parameter regimes.

\begin{figure*}
    \centering
  \includegraphics[width=\linewidth]{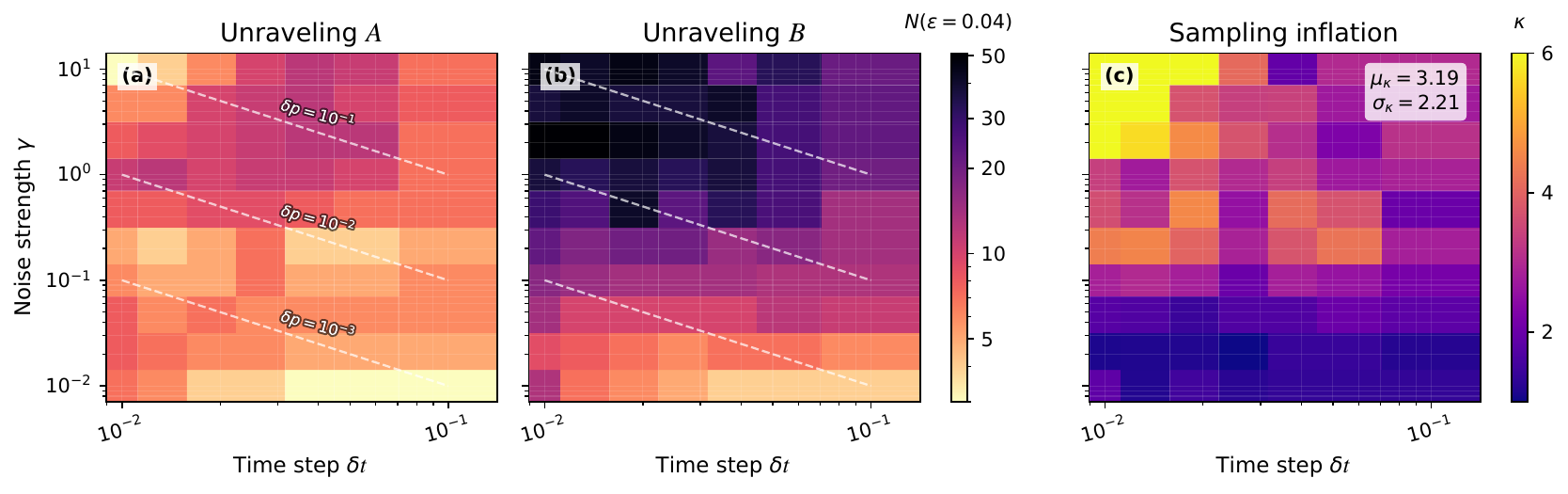}
    \caption{
    \textbf{Sampling requirements in a small-system exact-reference benchmark.}
    \textbf{(a,b)} Number of trajectories $N(\varepsilon=0.04)$ required for the trajectory-averaged
    correlator $\langle X_4 X_{5} \rangle$ to reach a fixed absolute deviation from exact reference evolution,
    shown as functions of noise strength $\gamma$ and time step $\delta t$ for a 10-site system evolved to
    $T=2$.
    Dashed contours indicate constant $\delta p=\gamma\delta t$.
    \textbf{(c)} Sampling inflation factor $\kappa = N_B/N_A$, quantifying the relative trajectory count
    required by unravelings A and B to reach the same reference-accuracy threshold.
    Because exact evolution is available for this small system, the extracted trajectory counts measure
    absolute error relative to the Lindblad reference; for large systems, the corresponding sampling cost is
    estimated from trajectory-to-trajectory variance through the standard error of the mean.
    }
    \label{fig:convergence}
\end{figure*}

\subsection{Trade-offs within individual trajectories}
The previous section showed that, for the TJM benchmark and unravelings considered, the observed bond
dimension reduction becomes less pronounced over the accessible finite-time system-size range.
We now fix the system size in this plateau-like regime and examine how trajectory-level complexity depends on
the interplay between physical noise strength $\gamma$ and numerical time step resolution $\delta t$.

Throughout this subsection we use time steps $\delta t \le 0.1$.
The second-order splitting and TDVP integration used in TJM have local errors
$\mathcal{O}(\delta t^3)$; the TDVP time-discretization error accumulates as
$\mathcal{O}(\delta t^2)$ at fixed evolution time~\cite{sander_TJM_2025}.

The effective jump probability per time step, $\delta p=\gamma\delta t$, is often used as a compact descriptor
of stochasticity.
However, it combines two distinct controls: the physical noise strength $\gamma$ and the numerical resolution
$\delta t$ at which stochastic events are sampled.
Fixing $\delta p$ therefore does not uniquely specify how Lindblad dynamics is resolved along individual
trajectories.

\cref{fig:gamma_dt} disentangles these effects by resolving the maximum bond dimension $\chi$ as a function of
$(\gamma,\delta t)$ for two physically equivalent unravelings.
Dashed contours indicate lines of constant $\delta p$.
All results are shown at fixed system size $L=65$ and final time $T=5$, where \cref{fig:Large_Scale} indicates
that the finite-time bond dimension trends have entered the plateau-like regime.

At fixed physical noise strength $\gamma$, decreasing the time step $\delta t$ systematically reduces $\chi$
for both unravelings.
Finer temporal resolution distributes stochastic events more evenly in time, interrupting coherent
entanglement growth and suppressing bond dimension accumulation.
Along contours of constant $\delta p$, however, $\chi$ still varies substantially with $\delta t$, showing
that jump probability alone does not determine trajectory-level entanglement complexity.
Coarser time steps sample stochastic events less frequently, allowing coherent evolution to generate larger
bond dimensions even when $\delta p$ is held fixed.

The two unravelings exhibit systematic quantitative differences across the $(\gamma,\delta t)$ plane,
summarized by the bond inflation factor $\alpha=\chi_A/\chi_B$ in \cref{fig:gamma_dt}(c).
Enhanced inflation, $\alpha\sim1.5$--$2.0$, appears at intermediate noise strengths and fine time steps, where
coherent entanglement growth competes with weak monitoring.
By contrast, $\alpha$ approaches unity both in the strongly monitored fine-time step regime and in the
coarse-time step regime, although for different reasons: in the former case, entanglement is rapidly
suppressed for both unravelings, while in the latter stochastic events are too coarsely resolved to generate
large unraveling-dependent differences.

These results show that discretization reshapes trajectory-level entanglement growth without changing the
underlying Lindblad generator.
Within the cost framework of Sec.~\ref{sec:cost_structure}, the resulting shifts in bond dimension define the
parameter-dependent inflation factor $\alpha(\gamma,\delta t)$.
Bond inflation alone, however, does not determine total simulation cost; trajectory sampling requirements
introduce an additional trade-off, captured by the sampling inflation factor $\kappa$.

\subsection{Sampling inflation at fixed physical accuracy}
\label{sec:sampling_inflation}

The previous sections characterized how physical and numerical choices reshape the bond dimension of
individual trajectories and thereby determine the bond inflation factor $\alpha$.
Total simulation cost, however, is governed by ensemble averages and therefore also depends on
trajectory-to-trajectory fluctuations.
We now quantify this second cost channel by extracting the sampling inflation factor $\kappa$ from exact
benchmarks.

\cref{fig:convergence} shows the
number of trajectories required to reach a fixed absolute error threshold $\epsilon\leq0.04$ across the full
time evolution in the central bond correlator $\langle X_4 X_{5} \rangle$ for a 10-site system with exact
reference evolution to $T=2$.
The deviation from this reference includes sampling error and any remaining time step or truncation bias.
The counts isolate sampling requirements only when these numerical biases are subdominant.

Both unravelings exhibit standard Monte Carlo convergence, $\varepsilon \propto {1}/{\sqrt{N}}$, in the
sampling-dominated regime.
The error pre-factors, however, depend strongly on $(\gamma,\delta t)$ and differ substantially between
unravelings.
These pre-factor differences define the sampling inflation factor $\kappa(\varepsilon)$ introduced in
Eq.~\eqref{eq:alpha_kappa_def}.

The resulting $\kappa(\gamma,\delta t)$ map in \cref{fig:convergence}\textbf{(c)} displays broad variation
across parameter space, with regions where one unraveling requires several times more trajectories than the
other to achieve identical accuracy.
Notably, the structure of $\kappa$ does not mirror that of the bond inflation factor $\alpha$: parameter
regions with reduced bond dimension can exhibit enhanced sampling requirements, and vice versa.
Sampling inflation therefore constitutes an independent cost channel that cannot be inferred from trajectory
entanglement alone.

Together $\alpha(\gamma,\delta t)$ and $\kappa(\gamma,\delta t)$ quantify two distinct forms of cost
redistribution: entanglement inflation within trajectories and variance inflation across trajectories.
In the next section, we combine these empirically extracted inflation factors with hardware constraints to
guide simulation design.

\section{Generalized cost extraction and hardware projections}
\label{sec:generalized_framework}

As an application of the cost analysis, pilot trajectories estimate the bond dimension and sampling inflation
factors $(\alpha,\kappa)$ for a specified simulation.
These estimates locate the comparison in the regime diagram and provide inputs for hardware projections.

We first describe the extraction protocol and then apply it to a Heisenberg-chain benchmark with depolarizing,
dephasing, and bit-flip noise.
This yields concrete hardware-aware decision maps illustrating how different noise channels populate the cost
plane.

\subsection{Protocol for extracting inflation factors}
\label{sec:extraction_protocol}
We consider two physically equivalent unravelings, $A$ and $B$, of the same Lindblad generator.
The objective is to determine the bond dimension inflation ratio $\alpha$ and the sampling inflation ratio
$\kappa$ for a fixed simulation scenario.

The inflation factors $(\alpha,\kappa)$ are scenario-dependent quantities.
For a fixed Hamiltonian, noise model, time step $\delta t$, final time $T$, initial state, observable, and
target statistical accuracy $\epsilon_{\mathrm{target}}$, they quantify how two physically equivalent
unravelings redistribute computational cost.
All quantities must therefore be evaluated under identical physical and numerical conditions.

For each unraveling $j\in\{A,B\}$, we perform a pilot trajectory simulation with a moderate number of
trajectories $N_{\mathrm{pilot}}$ and extract the maximum bond dimension $\chi_j$ reached during evolution.
Their ratio defines $\alpha$ for this simulation scenario.

To quantify sampling effort, we evaluate the variance of the chosen observable across trajectories at the
target time.
Let $\sigma_j$ denote the empirical standard deviation of unraveling $j$.
For a target standard error $\epsilon_{\mathrm{target}}$, the required trajectory count follows from the
central limit theorem as
\begin{equation}
N_j \approx \left(\frac{\sigma_j}{\epsilon_{\mathrm{target}}}\right)^2.
\end{equation} 
The corresponding ratio $N_B/N_A$ defines $\kappa$.
For a fixed simulation scenario, the standard error model at the target time gives
$\kappa\approx\sigma_B^2/\sigma_A^2$.
The same variance estimates can therefore be reused for a different target standard error if time step and
MPS biases remain subdominant.
This estimate concerns a single observation time.
The benchmark in Sec.~\ref{sec:sampling_inflation} uses absolute deviations from an exact reference across
the full time interval.

Pilot estimates carry finite-sample uncertainty and do not replace time step and MPS convergence checks.
They can be inserted into the cost model to locate the chosen scenario in the regime diagram.
The benchmarks reported here do not quantify net runtime savings after pilot overhead.

\begin{figure}
    \centering
    \includegraphics[width=\linewidth]{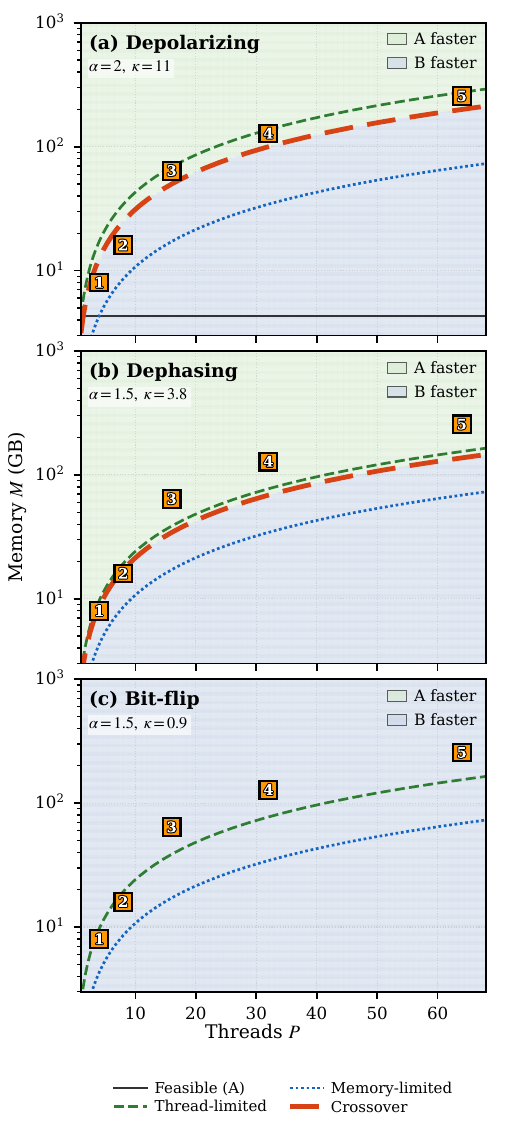}
    \caption{
    \textbf{Hardware-aware decision maps for Heisenberg noise channels.}
    Modeled runtime-minimizing regions in memory $M$ (GB) and workers $P$, using $L=16$ pilot estimates for
    (a) depolarizing $(\alpha,\kappa)=(2.0,11)$,
    (b) dephasing $(1.5,3.8)$,
    and (c) bit-flip $(1.5,0.9)$ noise.
    Green (blue) shading indicates where unraveling $A$ ($B$) yields lower modeled runtime.
    Solid, dashed, and dotted lines denote the feasibility threshold and memory/thread-limited boundaries,
    while the dash-dotted line marks the crossover $T_A=T_B$.
    Markers (1–5) represent budgets from 8 GB/4 threads to 256 GB/64 threads.
    The illustrative memory baseline is $M_B=L\chi_B^2\times16$ bytes for $L=1024$ and $\chi_B=256$.
    }
    \label{fig:mp_triptych}
\end{figure}

\subsection{Hardware-aware decision maps}
\label{sec:heisenberg_hardware}
To translate the extracted inflation factors into hardware-aware runtime estimates, we project each scenario
onto the hardware-aware decision geometry derived in ~\cref{sec:cost_structure}.
The decision maps in \cref{fig:mp_triptych} are constructed from the inflation factors
extracted for the Heisenberg chain with $L=16$, couplings $J=h=1$, noise strength $\gamma=0.1$, time step
$\delta t=0.1$, and final time $T=2$.
The observable is a local $Z$ expectation value, and trajectory counts are estimated to reach a target
standard error $\epsilon_{\mathrm{target}}=0.01$ using $N_{\mathrm{pilot}}=100$ pilot trajectories.

We consider three noise models, depolarizing ($X$, $Y$, $Z$ jumps), dephasing ($Z$), and bit-flip ($X$).
The Pauli jumps are unraveling A and their measurement-based analogs are unraveling B as done in
\cref{sec:benchmarking}.
The resulting $(\alpha,\kappa)$ values are therefore scenario-specific and serve here as concrete inputs to
the hardware projection.

We use an illustrative reference configuration with $L=1024$ and $\chi_B=256$ to display the hardware regimes.
Keeping the $L=16$ pilot ratios fixed in these maps is an assumption of the illustration, not evidence that
they remain unchanged at the larger size.
Assuming 16 bytes per complex number and unit implementation overhead, the baseline memory per trajectory of
unraveling $B$ is
\begin{equation}
M_B = L\,\chi_B^2 \times 16~\mathrm{bytes}.
\end{equation}

For fixed $(\alpha,\kappa)$, the modeled runtime ratio
\begin{equation}
\frac{T_A}{T_B} \sim \frac{\alpha^3}{\kappa}\,\frac{P_B}{P_A}
\end{equation}
induces a partition of the $(M,P)$ plane into regions where either unraveling~$A$ or $B$ has lower modeled
runtime.
Each panel of \cref{fig:mp_triptych} displays this decision map for one of the Heisenberg noise channels.
The solid line marks the feasibility threshold $M=\alpha^2 M_B$, dashed and dotted lines denote the onset of
thread- and memory-limited execution, and the dash-dotted line indicates the crossover $T_A=T_B$ when present.

To illustrate different resource budgets, we overlay representative hardware configurations
corresponding approximately to:
\begin{enumerate}
    \item embedded or edge devices ($M\sim8$~GB, $P\sim4$),
    \item laptops ($M\sim16$~GB, $P\sim8$),
    \item desktops ($M\sim64$~GB, $P\sim16$),
    \item small servers ($M\sim128$~GB, $P\sim32$),
    \item HPC nodes ($M\sim256$~GB, $P\sim64$).
\end{enumerate}
These markers illustrate how identical physical simulations may favor different unravelings depending on
available memory and parallelism.

The three noise channels therefore occupy distinct regions of the hardware decision geometry.
Using the reported $L=16$ depolarizing values $(\alpha,\kappa)=(2.0,11)$ gives
$\alpha^3=8<\kappa=11<32=\alpha^5$.
The scaling model predicts $T_A/T_B\approx8/11\approx0.73$ when both unravelings are worker-limited,
favoring~$A$, and $T_A/T_B\approx32/11\approx2.91$ when both are memory-limited, favoring~$B$.
The physical and numerical settings, observable, and target standard error are unchanged between these
limits.
These are model predictions using the reported pilot estimates, not measured runtime ratios.

Dephasing noise, $(1.5,3.8)$, remains in the trade-off quadrant but shifts the crossover toward lower memory,
enlarging the region where reduced bond dimension compensates sampling overhead.
In contrast, bit-flip noise, $(\alpha,\kappa)=(1.5,0.9)$, satisfies $\alpha>1$ and $\kappa<1$, so
unraveling~$B$ has both lower per-trajectory complexity and lower sampling cost and therefore enters the
strict-dominance regime.

Altogether, these cases illustrate that the analytical decision geometry is general, but the practical
unraveling choice depends sensitively on how a
given physical simulation populates the $(\alpha,\kappa)$ plane.

The runtime model should be interpreted as a resource-level decision criterion rather than a calibrated
wall-clock predictor.
The equations retain the dominant dependence on bond dimension, trajectory count, memory, and parallelism,
while omitting implementation-dependent prefactors from batching, scheduling, local tensor kernels,
linear-algebra backends, and communication overhead.
Such prefactors can be important whenever fixed overheads are comparable to the MPS evolution time.
The purpose of the hardware projections is therefore not to predict absolute runtimes, but to identify when
reduced per-trajectory complexity is expected to compensate, or fail to compensate, increased sampling effort.
Quantitative wall-clock optimization for a specific code path or architecture would require an additional
implementation-level calibration.

\Cref{tab:measured_resources} reports runtimes for a 16-site Heisenberg chain with depolarizing noise,
$\gamma=0.2$, $T=2$, $\delta t=0.005$, and the central $ZZ$ correlator.
The trajectory counts $(N_A,N_B)=(14,48)$ give comparable standard errors $(0.0203,0.0188)$.
At 416 MiB and $p\ge2$, memory admits one $A$ worker or two $B$ workers, and $B$ finishes 24--26\% sooner.
$A$ is faster in the remaining cells.
This comparison illustrates the effect of memory-constrained concurrency on unraveling choice.

\begin{table}[tb]
    \caption{
    Measured $A/B$ wall times in minutes.
    Columns give available workers $p$; memory limits apply to each complete ensemble.
    Bold entries mark the faster unraveling.
    }
    \label{tab:measured_resources}
    \centering
    \small
    \setlength{\tabcolsep}{2.8pt}
    \begin{tabular}{rcccc}
        \toprule
        \shortstack{Memory\\(MiB)} & $p=1$ & $p=2$ & $p=4$ & $p=6$ \\
        \midrule
        192  & $\mathbf{8.57}/13.03$ & $\mathbf{8.55}/12.82$ & $\mathbf{8.50}/12.81$ & $\mathbf{8.48}/12.85$ \\
        416  & $\mathbf{8.60}/13.01$ & $9.08/\mathbf{6.75}$ & $8.50/\mathbf{6.46}$ & $8.64/\mathbf{6.50}$ \\
        832  & $\mathbf{8.58}/12.89$ & $\mathbf{4.36}/6.53$  & $\mathbf{2.55}/3.30$  & $\mathbf{2.55}/2.74$ \\
        1536 & $\mathbf{8.59}/12.91$ & $\mathbf{4.35}/6.53$  & $\mathbf{2.55}/3.29$  & $\mathbf{2.00}/2.25$ \\
        \bottomrule
    \end{tabular}
\end{table}

\section{Unifying interpretation and outlook}
This work characterizes computational regimes created by the coupling between MPS trajectory storage and
parallel sampling.
Trajectory bond dimension affects both the work per sample and the number of samples that can fit concurrently
in memory.
Under the stated scaling assumptions, this changes the balance between trajectory complexity and sampling
effort.
Equivalent unravelings can therefore exchange their relative computational advantage between execution limited
by workers and execution limited by memory.
The inflation factors $(\alpha,\kappa)$ express these comparisons through relative bond dimensions and
required trajectory counts.

A central outcome is that reduced trajectory bond dimension does not automatically imply reduced total cost.
In the TJM/MCWF-style benchmarks studied here, environmental noise can suppress trajectory entanglement in
some regimes, but this reduction can weaken over the accessible finite time system size range or require finer
time step resolution.
In addition, an unraveling with lower per-trajectory complexity can require more trajectories to reach the
same statistical accuracy.
Bond dimension is therefore a useful diagnostic, but not a standalone complexity measure once discretization,
truncation, and sampling are taken into account.

As an application of the classification, the pilot protocol estimates bond dimension and sampling inflations
for a specified simulation.
Applying it to Heisenberg-chain benchmarks with different noise channels gives examples in different regions
of the $(\alpha,\kappa)$ plane, including both trade-off and strict-dominance regimes.
The boundaries follow from the scaling model, while the extracted values depend on the model, observable,
target accuracy, and simulation strategy.

The resulting hardware decision maps make unraveling choice an explicitly resource-dependent algorithmic
decision.
Physically equivalent stochastic representations may be favored or disfavored depending on memory capacity,
available parallelism, time step resolution, and sampling variance.
Thus, the preferred unraveling is not an intrinsic property of the Lindblad dynamics alone, but emerges from
the full computational setting.

The scope of this work is deliberately limited to exposing these cost channels rather than exhaustively
optimizing over all stochastic representations.
Other trajectory constructions may lead to different values of $(\alpha,\kappa)$ or different bond dimension
behavior.
This reinforces the main point: meaningful assessment of noisy quantum simulation requires explicit resource
accounting.
No single quantity, such as bond dimension, time step, noise strength, or trajectory count, provides a
universal complexity measure.
Instead, simulation strategies should be evaluated by how computational cost is distributed across
representation, sampling, and hardware resources.

\section*{Acknowledgments}
This work has been supported by the Munich Quantum Valley, which is supported by the Bavarian state government
with funds from the Hightech Agenda Bayern Plus (for which this work is the result of a collaboration
of two nodes).
It has also been supported by the Clusters of Excellence MATH+
and ML4Q.
It has additionally received funds from
the European Union under the Horizon Europe Programme as part of the European Research Council projects DA QC
(Grant Agreement 101001318) and DebuQC (Grant Agreement 101030988),
as well as Berlin Quantum (for which this work reflects joint work now involving the Weierstrass Institute
Berlin).
It has also received funding from the BMFTR (MuniQC-Atoms), the DFG (CRC 183, SPP 2514), and the Quantum
Flagship (PasQuans2, Millenion).
For MATH+, Millenion, SPP 2514, and the Munich Quantum Valley, this work is the result of a collaboration
involving more than one node each.

\bibliographystyle{apsrev4-2}
\bibliography{bib.bib}

\end{document}